\documentclass{ar-1col-S2O}
\usepackage[numbers,sort]{natbib}

\usepackage{url}
\usepackage{bm,bbold,amsmath}

\newcommand{\kB}{k_\mathrm{B}}
\renewcommand{\bf}[1]{\mathbf{#1}}
\jname{Xxxx. Xxx. Xxx. Xxx.}
\jvol{AA}
\jyear{YYYY}
\doi{10.1146/((please add article doi))}
\newcommand{\avg}[1] { \left < #1 \right > }

\begin{document}

\markboth{Singh et al.}{Variational path sampling of rare dynamical events}

\title{Variational path sampling
of rare dynamical events
}

\author{
Aditya N. Singh,$^{1,\dag}$ 
Avishek Das,$^{1,5,\ddag}$ 
David T. Limmer$^{1,2,3,4,*}$
\affil{$^1$Department of Chemistry, University, of California, Berkeley, USA}
\affil{$^2$Kavli Energy Nanoscience Institute, University, of California, Berkeley, USA}
\affil{$^3$Chemical Sciences Division, Lawrence Berkeley National Laboratory, Berkeley, CA, USA}
\affil{$^4$Material Sciences Division, Lawrence Berkeley National Laboratory, Berkeley, CA, USA}
\affil{$^5$Current address: AMOLF, Science Park 104, 1098 XG, Amsterdam, The Netherlands}
\affil{$^\dag$ansingh@berkeley.edu}
\affil{$^\ddag$avishek\_das@berkeley.edu}
\affil{$^*$dlimmer@berkeley.edu}}

\begin{abstract}
This article reviews the concepts and methods of variational path sampling. These methods allow computational studies of rare events in systems driven arbitrarily far from equilibrium. Based upon a statistical mechanics of trajectory space and leveraging the theory of large deviations, they provide a perspective with which dynamical phenomena can be studied with the same types of ensemble reweighting ideas that have been used for static equilibrium properties. Applications to chemical, material, and biophysical systems are highlighted.
 \end{abstract}

\maketitle


\section{Introduction}
In the past few years, a significant effort has been undertaken to extend the reach of modern molecular simulation tools into the realm of systems driven far from equilibrium~\cite{allen2009forward,dickson2010enhanced}. This effort is motivated in part by observation. It is increasingly possible to manipulate individual molecules~\cite{bustamante2003ten}, to build nanoscale devices subject to extreme driving forces~\cite{bocquet2010nanofluidics}, and observe active agents converting energy into directed motion~\cite{needleman2017active}. Such experiments drive molecular systems strongly enough that traditional near equilibrium approaches that form the core theory in physical chemistry are rendered inaccurate~\cite{fodor2022irreversibility,limmer2021large,ciliberto2017experiments}. At the same time, there have been abstract formal developments for nonequilibrium systems~\cite{limmer2024statistical}, including the formulation of stochastic thermodynamics~\cite{seifert2019stochastic} and dynamical large deviation theory~\cite{touchette2018introduction}. Stochastic thermodynamics has ushered in statements of symmetries of dynamical fluctuations~\cite{evans2002fluctuation,crooks1999entropy,jarzynski1997nonequilibrium}, and generalized response relationships~\cite{horowitz2020thermodynamic,baiesi2013update,speck2006restoring,gao2019nonlinear}. Large deviation theory has provided a language for probabilities of time dependent quantities~\cite{touchette2009large}. 

\begin{marginnote}
\entry{VPS}{Variational path sampling.}
\end{marginnote}
Motivated by a desire to bring to bear new theories of nonequilibrium states to detailed molecular systems, we and our co-workers have developed a general computational method for quantifying the likelihoods of dynamical events. The method requires no preconceived notion of the stationary configurational distribution, which is generically unknown when a system is driven from its equilibrium, Boltzmann state. 
Called \emph{variational path sampling} (VPS), it leverages the relationship between ensembles of trajectories conditioned on a specific dynamical event, and those driven to exhibit that event with high probability through the application of a control force~\cite{das2019variational,das2022direct}.  
This method generalizes old notions of Brownian bridges~\cite{doob1942brownian}, and brings them to bare on contemporary molecular problems. It lifts the importance sampling tools available for equilibrium molecular simulations into the study of nonequilibrium systems. The term \emph{variational} derives from a bound that the control force satisfies, a measure of the distance between the conditioned and driven trajectory ensembles, providing a basis for numerical approximation. \emph{Path sampling} refers both to the framework of the statistical mechanics of trajectories or paths, and to the ability to generate paths that lead to the rare dynamical event directly. 
We have demonstrated the construction of both approximate controllers and numerically exact ones, opening the way for computational studies of the dynamic pathways of chemical and physical transformations~\cite{singh2024splitting,singh2023variational,singh2024nonequilibrium}, and of rare fluctuations in driven and active systems~\cite{das2021reinforcement,grandpre2018current,grandpre2021entropy}.

\section{Rare events in and away from equilibrium} 
Typical fluctuations of molecular systems can be studied straight-forwardly by the direct integration of their equation of motion. Through such direct sampling, insights have been made into nonequilibrium systems and diverse phenomena discovered. However, in molecular systems rare events can be consequential.
For example, chemical reactions exhibit a strong separation of timescales--on average a molecule has to wait a long time before a reaction occurs relative to the fundamental timescales of molecular motion, and yet when it occurs it happens fleetingly, on the timescale of those typical motions~\cite{chandler1978statistical}. Therefore the likelihood of a reaction occurring at any given time is vanishingly small, yet once it  occurs, the properties of the substrate can be dramatically different. Analogously, collections of molecules in one phase can be transformed into another phase by the application of a suitable field. If that transition is between phases of different symmetry, the free energy along a distinguishing order parameter will exhibit two local minima, and the probability distribution along that coordinate will be bimodal. The existence of that second phase, and the coincident high susceptibility to the field, is encoded in the enhancement of probability far away from the typical behavior of the first phase, in states rarely if ever visited~\cite{binder1992monte}. 

Away from equilibrium, rare fluctuations play important roles in observable molecular behavior, though generically less is understood about their structure.  Chemical reactions and phase transitions occur under nonequilibrium driving with more diverse phenomenology than their equilibrium counterparts due to the breaking of time-reversal symmetry. We have found that rates of reactions are generically enhanced~\cite{kuznets2021dissipation} and the transitions forbidden at equilibrium can be accessed through the continuous dissipation of heat~\cite{singh2024nonequilibrium}.
We have been able to deduce the likelihoods and mechanisms of rare dynamical fluctuations including dynamic phase transitions~\cite{das2019variational,ray2018importance,ray2018exact,ray2019heat} using the VPS framework. 

\subsection{Path ensembles and stochastic dynamics} 
The perspective we adopt to study dynamical fluctuations is one in which the fundamental quantity is a stochastic trajectory, or path, of a system over time. By this we mean the time ordered sequence of configurations spanning a defined observation time, $\tau$. This perspective follows from work on dynamical fluctuations by Onsager, Machlup and Ruelle~\cite{onsager1953fluctuations,ruelle1999smooth}, and  on transition paths by Chandler and coworkers~\cite{bolhuis2002transition,chandler2010dynamics}. We will denote the configuration of a system at time $t$ as $\bf{x}_t$, which could include positions, momenta, or any other dynamical internal variable. A trajectory of a system will be denoted, $\bf{X}(\tau)=\{\bf{x}_t\}_{0\leq  t\leq \tau}$. Molecular systems obey Markovian equations of motion, which admit a factorized formulation of the probability of observing a trajectory, $P[\bf{X}(\tau)]$, given by
\begin{equation}
P[\bf{X}(\tau)] = \rho(\bf{x}_0)P[\bf{X}(\tau)|\bf{x}_0]
\end{equation}
the probability of observing the initial condition $\rho(\bf{x}_0)$ times the conditional probability of generating the subsequent trajectory from that initial condition, $P[\bf{X}(\tau)|\bf{x}_0]$. Moreover, $P[\bf{X}(\tau)|\bf{x}_0]$ itself can be further factorized over transition probabilities between configurations at arbitrary time displacements. 
 
To obtain a nonequilibrium steady state with persistent injection of energy, 
a system needs to be in contact with a bath that it can dissipate energy into. If the system is weakly coupled to a bath at constant temperature, the resulting equation of motion will take the following stochastic form
\begin{equation}
\dot{\bf{x}_t} = {\bm \mu}\bf{F}(\bf{x}_t) + {\bm \eta}_t
\end{equation}
where the dot represents time derivative, ${\bm \mu}\bf{F}(\bf{x}_t)$ is the time dependent flux of the configurations through phase space, and ${\bm \eta}_t$ is a Gaussian random variable with mean $\avg{{\bm \eta_t}}=0$, and covariance $\avg{{\bm \eta_t \otimes \bm \eta_{t'}}}=2 \kB T {\bm \mu} \delta(t-t')$ where the brackets denote thermal average while $\kB T$ is Boltzmann's constant times temperature~\cite{zwanzig2001nonequilibrium}. If $\bf{x}$ is a position, then ${\bm \mu}\bf{F}(\bf{x}_t)$ is a mobility matrix times a vector force. The force could include a gradient of a potential, $V(\bf{x}_t)$ as well as non-gradient drifts, $f(\bf{x}_t)$, that take the system away from equilibrium, $\bf{F}(\bf{x}_t)=-\nabla V(\bf{x}_t)+f(\bf{x}_t)$. The probability of a trajectory has an explicit form, written in terms of a stochastic action $\Gamma[\bf{X}(\tau)]$ with path integral measure $\mathcal{D}[\bf{X}(\tau)]$ satisfying
\begin{equation}
P[\bf{X}(\tau)|\bf{x}_0] = e^{-\Gamma[\bf{X}(\tau)]} \quad \quad 1 = \int \mathcal{D}[\bf{X}(\tau)] e^{-\Gamma[\bf{X}(\tau)]}
\end{equation}
\begin{marginnote}[]
\entry{Action}{The negative log likelihood of a trajectory, $\Gamma[\bf{X}(\tau)]$.}
\end{marginnote}
a normalization condition. The stochastic action can be written compactly as ~\cite{limmer2024statistical}
\begin{equation}
\Gamma[\bf{X}(\tau)] = \frac{1}{4\kB T }\int_0^\tau dt \left [ \dot{\bf{x}}_t - {\bm \mu}\bf{F}(\bf{x}_t) \right ] \cdot {\bm \mu}^{-1} \cdot \left [\dot{\bf{x}}_t -{\bm \mu} \bf{F}(\bf{x}_t)   \right ]
\end{equation}
where the quadratic form follows from the Gaussian noise.\footnote{There are subtleties associated with stochastic equations of motion, which arise when the noise depends on the state of the system. While not the case here, it is useful to state that the action above is in the  Ito convention, not Stratonovich where the action would include a gradient of the force~\cite{de2022path}.}

\begin{figure}
\includegraphics[width=16cm]{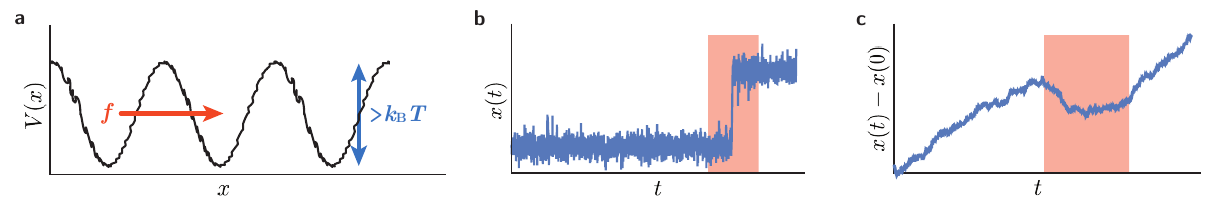}
\caption{Rare dynamical fluctuations away from equilibrium occur in systems like {\textbf a} where a particle evolves under both conservative, $-\nabla V(x)$ and non-conservative $f$ forces. Two examples of rare fluctuations include reactive transitions like the barrier crossing in {\textbf b} or a particle displacement in the direction opposite the force in {\textbf c}. Rare dynamical fluctuations are highlighted in red.}
\label{fig1}
\end{figure}

Given a sequence of configurations, the likelihood of a trajectory can thus be exactly computed. While the direct integration of the equation of motion will generate typical trajectories, atypical trajectories exist as well. Take for example a particle moving in a periodic potential under an external force like that pictured in Fig.~\ref{fig1}a. If the amplitude of the potential is large relative to $\kB T$, and large relative to the external force times the distance between the adjacent minima, 
the resultant dynamics of barrier crossing will look like that in Fig.~\ref{fig1}b, where the particle's position will dwell in one state for prolong periods of time, before fleetingly transitioning over the barrier into a new stable position. 
As a consequence, the likelihood of a barrier crossing trajectory of finite time is exceptionally low. 
Analogously, for a constant force propelling a particle in one direction, the second law requires that the particle displaces on average along that direction. 
If the force is large,  displacements against the applied force will be rare, and trajectories like that in Fig.~\ref{fig1}c, with prolonged domains of displacement against the force, will occur with low likelihoods. Below, we discuss quantifying both kinds of rare events.

\subsection{Rare events in fixed time} 
Transitions between collections of states, like those associated with chemical reactions and phase transitions, are examples of rare events that occur in finite time.
To discuss rates of transitions between two sets of states, it is convenient to introduce indicator functions that associate a given configuration to a given set. For example, for two sets of states denoted $A$ and $B$, we can define $h_A$ and $h_B$ such that
\begin{equation}
h_A[\bf{x}_t] = \begin{cases}  
1 \quad \bf{x}_t \in A \\
0 \quad \bf{x}_t \notin A \end{cases} \qquad h_B[\bf{x}_t] = \begin{cases}  
1 \quad \bf{x}_t \in B \\
0 \quad \bf{x}_t \notin B \end{cases}
\end{equation}
whose action on a configuration at a given time is to return 1 if a configuration is a member of that set and 0 otherwise. These functions act as filters, for example we can consider the trajectories of a system that start in set $A$ at time $t=0$, defined as a path partition function $Z_A$,
\begin{equation}
Z_{A} = \int d\bf{x}_0 h_A[\bf{x}_0]\rho[\bf{x}_0] = \avg{h_A[\bf{x}_0]}
\end{equation}
which by virtue of the condition acting only on the initial time does not depend on the subsequent path. Notice that $Z_A$ is equivalent to the average of the indicator function, or the probability of being in set $A$ in the steady state. If $A$ is made up of typical configurations, then there will be many paths that start in $A$, and the partition function will be large. Consider additionally filtering trajectories to be in $B$ at the end of the trajectory. The resultant partition function, $Z_{AB}(\tau)$, 
\begin{marginnote}[]
\entry{Path partition function}{An integral over a set of trajectories weighted by their probability, $Z$.}
\end{marginnote}
\begin{equation}
Z_{AB}(\tau) = \int d\bf{x}_0 \int \mathcal{D}[\bf{X}(\tau)]  h_A[\bf{x}_0]h_B[\bf{x}_\tau]\rho[\bf{x}_0]e^{-\Gamma[\bf{X}(\tau)]} =\avg{h_A[\bf{x}_0]h_B[\bf{x}_\tau]}
\end{equation}
now depends on the duration of the trajectory, $\tau$. The partition function can be recognized as an average, the joint probability that a trajectory starts in set $A$ and ends in set $B$. 

Just as their configurational analogues, ratios of path partition functions are related to physical quantities. The ratio $Z_{AB}(\tau)/Z_{A}$ has the probabilistic interpretation as the conditional probability of ending in $B$ at time $\tau$ given the system started in $A$ initially. The time rate of change of that probability, $k_\mathrm{AB}$, is thus
\begin{equation}\label{eq8}
k_\mathrm{AB} = \frac{d}{d \tau} \frac{Z_{AB}(\tau)}{Z_{A}} \sim \frac{1}{\tau} \frac{Z_{AB}(\tau)}{Z_{A}} 
\end{equation}
equivalent to a microscopic definition of the phenomenological rate constant~\cite{chandler1978statistical} for transitioning between metastable states~\cite{dellago1998transition}. Invoking the separation of timescales between the waiting time to transition and the characteristic time for the transition to occur over, it is expected that $Z_{AB}(\tau)$ will increase linearly in time, meaning that the time derivative is well-approximated by dividing by $\tau$.

\begin{marginnote}[]
\entry{Rate constant}{Probability per time of a reaction like $A$ goes to $B$, $k_{AB}$.}
\end{marginnote}

\subsection{Rare events in the long time limit} 
In complex systems, relevant dynamic variables are usually collective coordinates, and their fluctuations are encoded along their time-series. 
There are generically two different kinds of time-integrated, time-local observables~\cite{touchette2009large}, those that depend on the configuration denoted, $\mathcal{R}$, and those that depend on instantaneous changes to the configuration, $\mathcal{J}$. For trajectories of duration $\tau$, both can be defined as
\begin{equation}
\mathcal{R}[\bf{X}(\tau)] = \int_0^\tau dt \, r[{\bf{x}}_t] \, \qquad \mathcal{J}[\bf{X}(\tau)] = \int_0^\tau dt \, \bf{j}[{\bf{x}}_t]\cdot \dot{\bf{x}}_t  
\end{equation}
where $r[{\bf{x}}_t]$ and $\bf{j}[{\bf{x}}_t]$ are the time-local quantities associated with $\mathcal{R}$ and $\mathcal{J}[{\bf{x}}_t]$. The former is interpreted as an average of a configurational quantity over time. The latter is associated with currents of configurational quantities, and its integral is a generalized displacement.  

Just as we considered trajectories filtered for transitioning between two sets of states, we can consider the ensemble of trajectories conditioned on a value of a specific parameter. Take $\mathcal{O}[\bf{X}(\tau)]$ to be some generic time integrated observable. The partition function for trajectories conditioned on a specific value of $\mathcal{O}[\bf{X}(\tau)]$, $Z_\mathcal{O}$, is computed from
\begin{equation}
Z_\mathcal{O} = \int d\bf{x}_0 \int \mathcal{D}[\bf{X}(\tau)]  \delta \left \{ \mathcal{O}[\bf{X}(\tau)]-{\mathcal{O}} \right \}\rho[\bf{x}_0]e^{-\Gamma[\bf{X}(\tau)]} = \avg{ \delta \left \{ \mathcal{O}[\bf{X}(\tau)]-{\mathcal{O}} \right \}}
\end{equation}
which is recognized as the average of the delta function, or equivalently the probability that trajectories of duration $\tau$ exhibit $\mathcal{O}$. In the limit of long observation times for time integrated quantities, provided a finite correlation time for fluctuations in $\mathcal{O}$, the partition function or probability takes on a simplified form
\begin{equation}\label{eqratef}
I(\mathcal{O}/\tau) =\lim_{\tau \rightarrow \infty} -\frac{1}{\tau}\ln Z_\mathcal{O}
\end{equation}
where $I(\mathcal{O}/\tau)$ is a so-called rate function from large deviation theory~\cite{touchette2018introduction}, a negative log-likelihood independent of the observation time. The rate function characterizes the spectrum of fluctuations of time-integrated observables. Typical fluctuations encoded in $I(\mathcal{O}/\tau)$ portray expectations from the central limit theorem and its concomitant Gaussian fluctuations. Encoded in the extreme fluctuations, those trajectories consistent with $Z_\mathcal{O}$ for $\mathcal{O}$ far away from the mean, are molecular mechanisms for improbable dynamical fluctuations. Signatures of collective effects like nonequilibrium phase transitions can show up as significant enhancement of rare fluctuations relative to priors informed from the central limit theorem, or neglect of correlations. 
Moreover, the rate function and its dependence on the driving forces contain information about the response of the system to changing driving forces, through generalized fluctuation-dissipation relationships~\cite{speck2016thermodynamic,gao2019nonlinear,lesnicki2020field,gao2017transport,ray2019heat}. 

\begin{marginnote}[]
\entry{Rate function}{The time intensive negative log-likelihood of a time integrated observable, $I(\mathcal{O}/\tau)$.}
\end{marginnote}

\section{Path ensemble reweighting}  
The scarcity of rare but important events renders them very difficult to study computationally. In equilibrium systems, knowledge of the stationary distribution enables significant simplification. Enhanced sampling tools, from traditional umbrella sampling to current artificial intelligence augmented methods, have rendered the sampling of rare configurational fluctuations and the evaluation of their likelihoods straightforward in all but the most pernicious cases~\cite{binder2012monte,mehdi2024enhanced}. These methods increase the likelihood of specific configurations by adding an external potential, and account for the change in the probability by leveraging Boltzmann statistics. Optimal ways of reweighting configurational probabilities, decoding the resulting information, and doing so with high accuracy models is still difficult for some systems and important developments are still being made along these lines~\cite{henin2022enhanced}. 

The study of dynamical quantities is generally more challenging. Adding potentials to enhance the sampling of rare configurations necessarily alters the dynamics with which a system evolves through those configurations, often in a manner that is difficult to interpret. In some cases, theory can connect configurational fluctuations to dynamical ones, as is done with transition state theory 
~\cite{peters2017reaction}. However, these relationships typically require the system to be in an equilibrium. Path sampling approaches provide a means of directly importance sampling dynamical quantities~\cite{bolhuis2002transition}. While most efficient for the study of transitions accompanying a chemical reaction, path sampling methods have been extended to diffusive events~\cite{guttenberg2012steered,gingrich2015preserving,grunwald2008precision}. The efficiency of traditional path sampling relies on microscopic reversibility, which is violated in systems away from equilibrium. Variations~\cite{crooks2001efficient,buijsman2020transition,ray2018importance} and alternatives~\cite{allen2009forward,dickson2010enhanced,zuckerman2017weighted} have been proposed to evaluate nonequilibrium averages and rates of rare events. These strategies stratify the rare fluctuation into a sequence of less rare events that can be concatenated. Only recently have methods like VPS been developed that importance sample dynamical quantities by directly enhancing their likelihood through a change to their equation of motion. VPS does this through path ensemble reweighting, in close analogy to traditional methods of configurational enhanced sampling. 

\subsection{Reweighting in equilibrium} 
To illustrate how path ensemble reweighting works, it is useful to briefly review importance sampling and reweighting for configurational statistics in equilibrium. In equilibrium, at fixed temperature, volume and number of particles,  the steady state distribution is given by a Boltzmann form dependent only on the internal energy of the system. If the system's Hamiltonian is $\mathcal{H}_0(\bf{x})$ then the resultant distribution is
\begin{equation}
    \rho_0(\bf{x})=\frac{1}{Q_0}e^{-\mathcal{H}_0(\bf{x})/\kB T} \qquad Q_0 = \int d\bf{x}\, e^{-\mathcal{H}_0(\bf{x})/\kB T}
\end{equation}
where $Q_0$ is the canonical partition function, with $d\bf{x}$ a suitably chosen measure, often made dimensionless by introducing factors of Planck's constant. The likelihood of observing a given configuration of the system under a different Hamiltonian, $\mathcal{H}_1(\bf{x})$, can be straightforwardly related to the first distribution,
\begin{equation}
\rho_0(\bf{x})=\rho_1(\bf{x}) \frac{Q_1}{Q_0}e^{-\Delta \mathcal{H}(\bf{x})/\kB T} \qquad Q_0 = Q_1 \avg{e^{-\Delta \mathcal{H}(\bf{x})/\kB T}}_1
\end{equation}
where $\Delta \mathcal{H}=\mathcal{H}_0-\mathcal{H}_1$ and the average is taken under the new Hamiltonian. The probability of a configuration in one ensemble, or under a given Hamiltonian, is related to the probability under a different Hamiltonian times a reweighting factor. Given an order parameter, or putative reaction coordinate, $\mathcal{O}$, whose statistics are of interest, the marginal distribution can be computed by averaging Dirac's delta function, $p_0(\mathcal{O})=\avg{\delta[\mathcal{O}-\mathcal{O}(\bf{x})]}_0$, in the original ensemble. In most cases there will be regions of $p_0(\mathcal{O})$  in which the system is unlikely to visit, such as transition states or distinct metastable minima. Within equilibrium importance sampling, one can add a potential that is a function of $\mathcal{O}$, such that $\Delta \mathcal{H}=U(\mathcal{O})$  to enhance the likelihood of a given value of $\mathcal{O}$~\cite{frenkel2023understanding}. Through the reweighting principle above, the distribution of order parameter values generated with the extra potential is related to the original ensemble as 
\begin{equation}
p_0(\mathcal{O})=p_1(\mathcal{O})\frac{Q_1}{Q_0}e^{-U(\mathcal{O})/\kB T} \qquad Q_0 = Q_1 \avg{e^{-U(\mathcal{O})/\kB T}}_1
\end{equation}
providing access to the free energy, or reversible work, to change the parameter, $\Delta F=- \kB T \ln p_0(\mathcal{O})$, in the original ensemble.

\subsection{Conditioned, tilted, and driven ensembles} 
In analogy to equilibrium, reweighting trajectory ensembles allows  for rare fluctuations to be observed with higher probabilities, admitting studies on their mechanisms and origins. Two important distinctions exist for reweighting paths. First, the central limit theorem dictates that distributions of sums of random variables sharpen around their mean. Distributions of observables that are extensive in particle number and that are integrated over time are consequently very sharply peaked around their typical values, or equivalently that fluctuations away from the typical value are less likely. This inevitably means that more computational power is needed to sample them than their configurational counterparts. Second, equilibrium configurational properties are independent on the details of the equations of motion used to sampling them. By construction, dynamical properties lose this universality. This can render their interpretation difficult, and care must be taken in order to pose the relevant equation of motion. 
The mathematics of reweighting Markovian processes largely go back to old work by Doob and Girsanov~\cite{doob1984classical,girsanov1960transforming}. 

\begin{figure}[t]
\includegraphics[width=15.5cm]{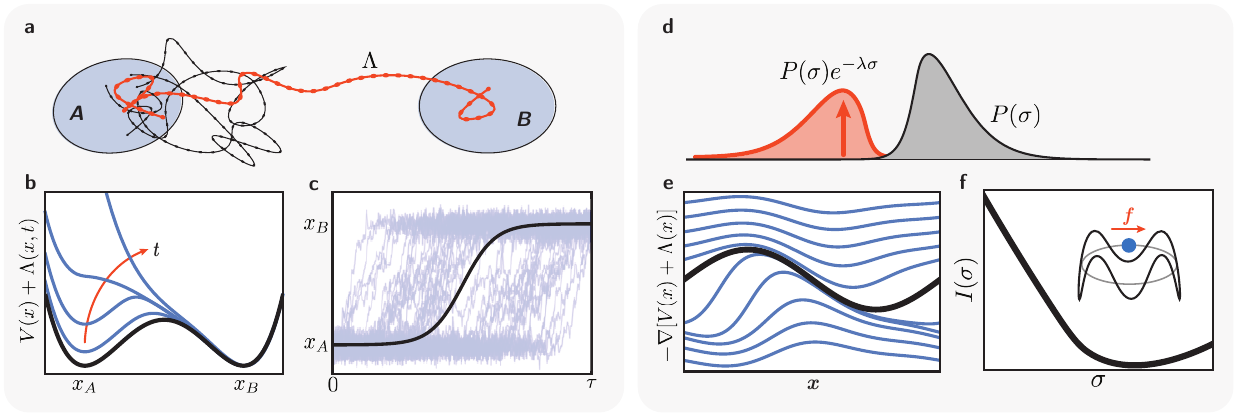}
\caption{Illustration of conditioned, tilted and driven trajectory ensembles. {\textbf a} A typical trajectory (black) evolving in a metastable system between states A and B is rarely reactive (red) unless driven by a control force $\Lambda$. {\textbf b} An example bistable system and time dependent potential to affect transitions. {\textbf c} Reactive trajectories (blue) and their mean (black) with evident time-translational invariance. {\textbf d} Time integrated quantities like the entropy production $\sigma$ exhibit typical fluctuations (black) and can be biased to exhibit rare fluctuations by tilting with a parameter $\lambda$ (red). {\textbf e} Control forces (blue) on top of the conservative force (black) that generate rare entropy fluctuations for a system in {\textbf f} of a particle driven around a ring whose rate function has exponentially small probabilities for negative values due to Gallavotti-Cohen symmetry \cite{gallavotti1995dynamical}}
\label{fig2}
\end{figure}
\begin{marginnote}[]
\entry{Path ensemble}{Collection of trajectories with a common constraint.}
\end{marginnote}

We can construct a path ensemble conditioned on exhibiting a dynamical event, $\mathcal{O}$, by manipulating the original path probability
\begin{equation}
P_\mathcal{O}[\bf{X}(\tau)] = \frac{1}{Z_\mathcal{O}} \delta \left \{ \mathcal{O}[\bf{X}(\tau)]-{\mathcal{O}} \right \}\rho[\bf{x}_0]e^{-\Gamma[\bf{X}(\tau)]} 
\end{equation}
where the delta function provides the filtering of trajectories and $Z_\mathcal{O}$ renormalizes the distribution. 
Choosing $\mathcal{O}[\bf{X}(\tau)]=h_A[\bf{x}_0]h_B[\bf{x}_\tau]$, generates the reactive path ensemble, $P_{AB}[\bf{X}(\tau)]$ with transition path partition function, $Z_{AB}(\tau)$~\cite{dellago1999calculation}. For example in Fig.~\ref{fig2}a, for a system that exhibits two metastable basins, denoted $A$ and $B$, most trajectories started in $A$ will remain, or will only exhibit brief excursions out of that basin, with high likelihood of returning. Within the conditioned path ensemble, all trajectories react, albeit with varying probabilities. Rather than sampling from $P_{\mathcal{O}}[\bf{X}(\tau)]$, inspired by traditional equilibrium techniques, one could alternatively alter the system in order to make improbable trajectories typical. For example, one could add an external time dependent potential that drives the system to a rare event, ${\Lambda}(\bf{x}_t,t)$, with an equation of motion 
\begin{marginnote}[]
\entry{Transition path ensemble}{Collection of trajectories that react.}
\end{marginnote}
\begin{equation}\label{eqeom2}
\dot{\bf{x}_t} = {\bm \mu}\bf{F}(\bf{x}_t)-{\bm \mu}\nabla {\Lambda}(\bf{x}_t,t) + {\bm \eta}_t
\end{equation}
as illustrated for a reactive conditioning in Fig.~\ref{fig2}b. The consequence of the added potential can be understood quantitatively by relating the conditioned path probability to a conditioned path probability under the external driving, denoted $P_{\mathcal{O},\Lambda}[\bf{X}(\tau)]$, assuming the initial condition is the same,
\begin{equation}\label{eq:19}
P_{\mathcal{O}}[\bf{X}(\tau)] = P_{\mathcal{O},\Lambda}[\bf{X}(\tau)] \frac{1}{Z_{\mathcal{O}}(\tau)} e^{- \Delta \Gamma[\bf{X}(\tau)]}
\end{equation}
where the path partition functions are related by
\begin{equation}\label{eqexpav1}
Z_{\mathcal{O}}(\tau) = Z_{\mathcal{O},\Lambda}(\tau) \avg{e^{- \Delta \Gamma[\bf{X}(\tau)]}}_{\mathcal{O},\Lambda}
\end{equation}
and $\Delta \Gamma$ is the change in stochastic action~\cite{kuznets2021dissipation}. For the added potential to the stochastic equation of motion above, the change in action is
\begin{equation}
\Delta \Gamma = -\frac{1}{4 \kB T} \int_0^\tau dt \, \nabla \Lambda(\bf{x}_t,t) \cdot \left [ 2\dot{\bf{x}}_t - 2{\bm \mu}\bf{F}(\bf{x}_t) + {\bm \mu} \nabla \Lambda(\bf{x}_t,t) \right ] 
\end{equation}
which depends on the details of the individual trajectories. These reweighting statements are valid for any arbitrary added force. We refer to the collection of trajectories evolved with the added force as the driven ensemble of trajectories, and Eq.~\ref{eq:19} is a relationship between the conditioned path ensemble and the driven one.
\begin{marginnote}[]
\entry{Driven path ensemble}{Collection of trajectories that react due to an extra, external force.}
\end{marginnote}

Using the previous relation between the rate constant and the ratio of path partition functions, Eq.~\ref{eq8}, we can define the rate in the driven ensemble analogously,
\begin{equation}
k_\mathrm{AB,\Lambda} = \frac{1}{\tau} \frac{Z_{AB,\Lambda}(\tau)}{Z_{A}}\sim \frac{1}{\tau}
\end{equation}
where if the driving force is chosen such that it enforces all trajectories to be reactive, the ratio of partition functions will be one, or the rate in the driven ensemble is just the inverse observation time. Using the reweighting relation, we can relate the rates in the driven and conditioned ensemble in the limit that the added force ensures reactivity,
\begin{equation}\label{eqvpsk}
k_\mathrm{AB} \tau =    \avg{e^{-\Delta \Gamma[\bf{X}(\tau)]}}_\Lambda \geq  e^{-\avg{\Delta \Gamma[\bf{X}(\tau)]}_\Lambda}
\end{equation}
where the first equality is true for all added potentials that enforce reactivity and the inequality follows from the convexity of the exponential and Jensen's inequality~\cite{das2022direct}. An equivalent expression relates the rate to the average change in action in the conditioned ensemble without $\Lambda$, $\ln k_\mathrm{AB} \tau \leq \langle \Delta \Gamma \rangle_{AB}$. While the equality relates the rate constant in the original system to an average in the driven ensemble is exact, it is not of numerical use as exponential averages are difficult to converge. The inequality is conversely easy to evaluate as it is a simple average, but for an arbitrary  potential it may provide a poor estimate of the rate. This inequality forms the basis of the way VPS evaluates rate constants, by variationally optimizing $\Lambda$ to saturate the bound. The inequality relates the rate in the reference ensemble to the mean change in action, or equivalently since the action is the log likelihood, the rate is given by the Kullback-Leibler divergence between the conditioned and driven ensembles. This observation provides an interpretation for the saturation of the bound, as given by a specific time dependent potential that renders the subsequent reactive trajectories statistically indistinguishable from those of the conditioned ensemble. In Fig.~\ref{fig2}c, the time-translational invariance of a steady state would ensure that a reaction would be equally likely to happen at any point over the $\tau$ observation time. Driven trajectories should similarly react equally likely anywhere in that time window.  

An alternative to the conditioned ensembles are so-called tilted ensembles, which have been formulated most often for time-integrated observables ~\cite{chandler2010dynamics,garrahan2009first,hedges2009dynamic,hurtado2011spontaneous,baek2017dynamical,vaikuntanathan2014dynamic,garrahan2010thermodynamics}. This ensemble is inspired by the Laplace transform structure of equilibrium statistical mechanics and weights paths exponentially in proportion to the observable of interest times a counting parameter, denoted here as $\lambda$. Shown in Fig.~\ref{fig2}d, the addition of this weighting factor, with specifically chosen values of $\lambda$, can enhance the likelihood of otherwise rare fluctuations of $\mathcal{O}$. The resultant distribution of trajectories in the tilted ensemble has probability
\begin{marginnote}[]
\entry{Tilted path ensemble}{Collection of trajectories reweighted with a linear bias.}
\end{marginnote}
\begin{equation}
P_\lambda[\bf{X}(\tau)] = \frac{1}{Z_\lambda}\rho[\bf{x}_0]e^{-\Gamma[\bf{X}(\tau)]} e^{-\lambda \mathcal{O}[\bf{X}(\tau)]}
\end{equation}
where in the long time limit the initial condition distribution is irrelevant, and
\begin{equation}\label{scgf}
Z_\lambda = \int d\bf{x}_0 \int \mathcal{D}[\bf{X}(\tau)]   \rho[\bf{x}_0]e^{-\Gamma[\bf{X}(\tau)]} e^{-\lambda \mathcal{O}[\bf{X}(\tau)]}=\avg{e^{-\lambda \mathcal{O}}}
\end{equation}
is the tilted partition function, equal to the moment generating function, or average exponential of the tilting factor $\lambda \mathcal{O}$. Equivalently, $-(\ln Z_\lambda)/\tau$ is a scaled cumulant generating function. The Laplace structure of the path distribution manifests in the long time limit as a Legendre-transform structure that relates the scaled cumulant generating function to the original rate function,
\begin{equation}\label{eqLFT}
I(\mathcal{O}/\tau) = \frac{1}{\tau} \max_\lambda \left [ \lambda  \mathcal{O} +\ln Z_\lambda \right ]
\end{equation}
valid when $\tau$ is much longer than the correlation time of $\mathcal{O}$. Working within the tilted ensemble provides equivalent access to the statistics of time integrated observables. 

\begin{marginnote}[]
\entry{Scaled cumulant generating function}{Negative logarithm of the time intensive Laplace transform of a probability distribution, $-(\ln Z_\lambda)/\tau$.}
\end{marginnote}

There are methods to compute $\ln Z_\lambda$ and sample $P_\lambda[\bf{X}(\tau)]$, which have been applied to molecular systems~\cite{jack2006space,hedges2009dynamic,giardina2011simulating}.
Like methods available to compute rates away from equilibrium, these methods stratify the sampling of $Z_\lambda$ but do not make the fluctuations that contribute most at each value of $\lambda$ more probable. An alternative is to drive the system directly to the rare value of $\mathcal{O}$. For example an additional potential, $\bf{\Lambda}(\bf{x}_t)$ could be added to the equation of motion as in Eq.~\ref{eqeom2}, 
like those drifts shown in Fig.~\ref{fig2}e to probe rare particle displacements. With judiciously chosen $\bf{\Lambda}(\bf{x}_t)$, the same values of $\mathcal{O}$ could be sampled as would for a given value of $\lambda$.
The resulting distribution of trajectories, a driven ensemble, is related to the original distribution, $P_\lambda[\bf{X}(\tau)]$, as
\begin{equation}
P_\lambda[\bf{X}(\tau)] = P_{\lambda,\Lambda}[\bf{X}(\tau)]  \frac{1}{Z_{\lambda}} e^{-\Delta \Gamma[\bf{X}(\tau)]}
\end{equation}
where the partition function is evaluatable as
\begin{equation}\label{eqvpsldf}
Z_{\lambda} =  \avg{e^{-\lambda \mathcal{O}[\bf{X}(\tau)]- \Delta \Gamma[\bf{X}(\tau)]}}_\Lambda \geq e^{-\lambda \avg{ \mathcal{O}[\bf{X}(\tau)]}_\Lambda- \avg{\Delta \Gamma[\bf{X}(\tau)]}_\Lambda }
\end{equation}
since $P_{\lambda,\Lambda}[\bf{X}(\tau)]$ is a normalized distribution so that its partition function is equal to 1~\cite{das2019variational}. The evaluation of $\ln Z_{\lambda}$ or equivalently $I(\mathcal{O}/\tau)$ is equivalent to averaging the exponential in the first equality under the driving dynamics, providing access to rare time integrated quantities away from equilibrium, an example of which is in Fig.~\ref{fig2}f. The second inequality acts similar to Eq.~\ref{eqvpsk} to provide a variational bound to determine $\ln Z_{\lambda}$. Analogously, VPS algorithms can be developed to variationally approach the calculation of $I(\mathcal{O}/\tau)$.

\subsection{Relation to stochastic control}  
The variational inequalities in Eq.~\ref{eqvpsk} and \ref{eqvpsldf} can be saturated with unique added forces known from previous work in the literature on stochastic processes and have close connections to theories of stochastic control. For this reason we refer to the added drift that enters into the reweighting expressions as the control force, though it need not be a force in the Newtonian sense. At saturation both equalities follow from a so-called generalized Doob transform~\cite{chetrite2015variational}. For the rate constant problem in Eq.~\ref{eqvpsk} the optimal control force, $-\nabla \Lambda^*$, follows from the solution of a Cole-Hopf transformed Backward Kolmogorov equation~\cite{das2022direct},
\begin{marginnote}[]
\entry{Optimal control force}{A force that renders rare fluctuations typical in a manner indistinguishable from an unforced spontaneous fluctuation, $-\nabla \Lambda^*$.}
\end{marginnote}
\begin{equation}\label{eq:bke}
\frac{\partial}{\partial t} \Lambda^* = - {\bm \mu} \bf{F} \cdot \nabla \Lambda^* -\kB T \nabla \cdot ({\bm \mu} \nabla \Lambda^* ) + \nabla \Lambda^* \cdot\left ( {\bm \mu} \nabla \Lambda^* \right ) /2
\end{equation}
with boundary conditions applied at the final time, $\Lambda (\bf{x}\in B,t=\tau) =0$.  The boundary condition enforces that the trajectories react. In the absence of a force in the reference system, this equation is that for a Brownian-Bridge first solved by Doob~\cite{doob1957conditional}. Orland and coworkers generalized this to barrier crossing events in equilibrium systems, and employed analytical approximations to it as a means of generating transition paths, though did not recognize the connection to the evaluation of rates~\cite{orland2011generating,koehl2022sampling,majumdar2015effective}. Subsequent work~\cite{hartmann2012efficient,hartmann2013characterization,zhang2014applications}, building off of the transition path theory formalism~\cite{vanden2010transition}, recognized the stationary solution of the optimal control problem in Eq.~\ref{eq:bke} as isomorphic to that of the committor, or splitting probability, $\bar{q}_B(\bf{x}) = \exp[- \Lambda^*(\bf{x}) {\bm\mu}^{-1}/2 \kB T]$. We refer to the time-dependent solution of the Backward Kolmogorov equation as the time-dependent committor $q_B(\mathbf{x},t)$ \cite{das2022direct,majumdar2015effective}, which is related to the optimal controller through $\Lambda^*(\mathbf{x},t) = -2 \kB T \ln q_B(\mathbf{x},t)$~\cite{singh2024splitting}.

\begin{marginnote}[]
\entry{Committor}{The probability of reaching state $B$ before state $A$ starting in configuration $\bf{x}$, $\bar{q}_B(\bf{x})$.}
\end{marginnote}

For the evaluation of large deviation functions for observables of the form $\mathcal{O} = \mathcal{J}+\mathcal{R}$, the optimal control force satisfies a Hamilton-Jacobi-Bellman equation~\cite{chetrite2015variational,rose2021reinforcement},
\begin{align}\label{eqHJB}
&\frac{1}{2\kB T} \left [ \nabla \Lambda^* \cdot\left ( {\bm \mu} \nabla \Lambda^* \right )- 2({\bm \mu} \bf{F}) \cdot \nabla \Lambda^*\right ] - \nabla \cdot ({\bm \mu} \nabla \Lambda^* ) \\
&=2\lambda r+2\psi(\lambda)-2\lambda^{2}\kB T\mathbf{j}\cdot(\bm{\mu}\mathbf{j})-\lambda\Big\{ \mathbf{j}\cdot(\bm{\mu}\nabla\Lambda^* )+(\bm{\mu}\mathbf{j})\cdot\nabla\Lambda^* -2(\bm{\mu}\mathbf{F})\cdot\mathbf{j}-2 \kB T\nabla\cdot(\bm{\mu}\mathbf{j})\Big\} \nonumber
\end{align}
where $\psi(\lambda)=(\ln Z_\lambda)/\tau$ is the scaled cumulant generating function. 
The optimal controller for generating trajectories biased on large deviations, $P_{\lambda}[\mathbf{X}(\tau)]$, is related to the eigenfunction of a tilted Fokker-Planck equation~\cite{chetrite2015nonequilibrium}. There is however one additional subtlety. If $\mathcal{O}$ includes a current-type variable, $\mathcal{J}$, then the optimal control force cannot be written as a gradient. It contains a gradient part from the solution of the above equation, and an additional non-gradient part, such that $\nabla \Lambda^* \rightarrow 2 \kB T \lambda \mathbf{j}  +\nabla \Lambda^*$. The long time limit in the tilting of the trajectory ensemble renders the optimal control force time-independent. There has been significant work developing the optimal control theory for processes conditioned on large deviations of integrated observables, beginning with its identification~\cite{jack2015effective,chetrite2015nonequilibrium} and following with a variety of numerical approaches~\cite{ray2018exact,nemoto2014computation,nemoto2016population}. Efficient basis sets like tensor networks have enabled the direct numerical solution of discrete analogies to Eq.~\ref{eqHJB} for lattice models~\cite{causer2023optimal,banuls2019using,helms2019dynamical}. Feedback control and alternative variational approaches have also been considered and applied to systems in the continuous space~\cite{yan2022learning,nemoto2019optimizing}.

\section{Variational path sampling} 
The goal of VPS is to compute the statistics and mechanisms of rare fluctuations by learning the optimal controllers that make the rare events \textit{typical}. This is done by the parameterization of the control force $-\nabla \Lambda$ through the prescription of an ansatz and minimization of the gradients of the variational cost functions on the right-hand side of Eqs. \ref{eqvpsk} and \ref{eqvpsldf} with respect to the parameters.
 At heart of VPS is the estimation of these gradients efficiently from simulated trajectories given a functional form of $\Lambda$. The optimized forces are then used to obtain estimates of the likelihood of the rare fluctuation, and can be used to generate the ensemble of trajectories that achieve the fluctuation, from which mechanistic insight can be gained.

\subsection{Ansatz for the control force}
Since the optimal controllers are related to solutions of the Fokker-Planck and Backward Kolmogorov equations, they formally span the whole dimensional space of the system. Nevertheless, the  parameterization of forces along a low-rank set of descriptors can be both efficient and accurate.
Once a set of descriptors are chosen, one has to choose an ansatz for parameterization of the forces. This choice strongly depends on the physical symmetries of the system and the nature of the rare fluctuation. Since the optimal force generally respects the highest symmetries common to both the system dynamics and the observable, particular advantage can be gained by choosing an ansatz that also obeys the same symmetries. For example, in a spatially periodic system, sampling rare values of a current that does not break that symmetry is achieved with an optimal controller that is also periodic. In many-particle systems with a spatial translation symmetry, sampling rare values of dynamical activity without reference to absolute positions corresponds to a controller that is also translation-invariant. The importance of these symmetries have analogously been a guiding principle for development of neural-network based forcefields for molecular dynamics~\cite{kocer2022neural,keith2021combining}.

\begin{figure}
\includegraphics[width=16cm]{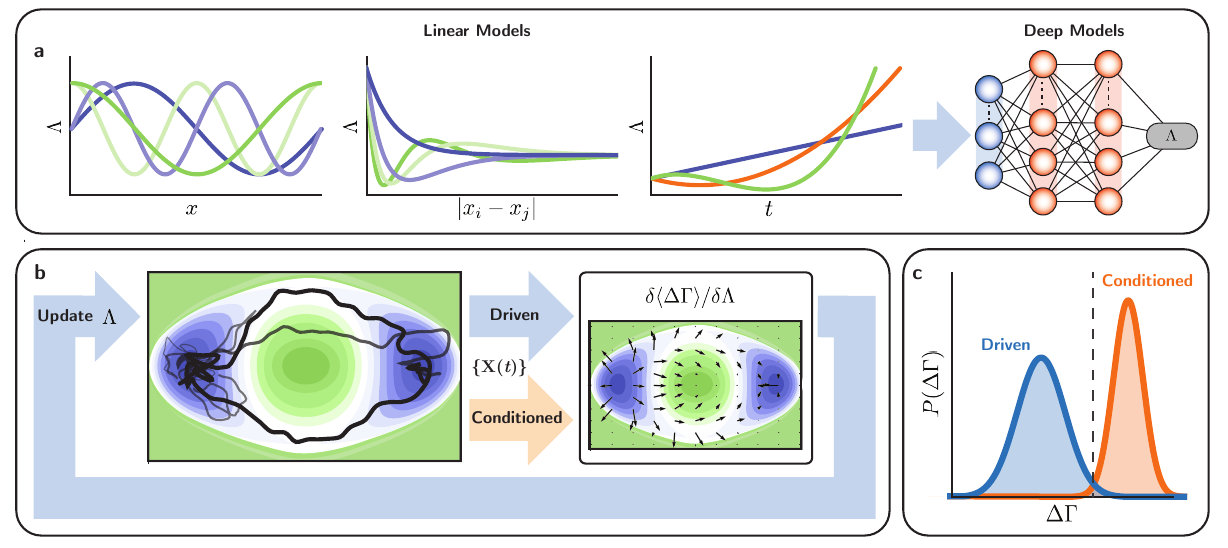}
\caption{Schematic of the Variational Path Sampling algorithm. {\textbf a} A representation for the control force is  determined by defining descriptors in space and time. These are spanned by basis functions, for example Fourier (left), Laguerre (middle) or Jacobi (right) functions. The force can be represented by linear combinations of these descriptors or non-linear models like deep feed-forward neural networks. {\textbf b} Optimization of the control force follows from either one-shot fitting (orange) or iterative learning procedures (blue). The rare event being sampled is a barrier crossing transition from the left to the right well on the potential energy surface. The learning procedure requires several iterations to obtain both the rare trajectories as well as optimization of the control force. The black lines represent reactive trajectories and the increasing opaqueness highlights the decrease in Kullback–Leibler divergence as the optimization proceeds. {\textbf c} Estimators can be evaluated based on either the conditioned or driven ensembles, and their accuracy depends on the spread of the distribution of $\Delta \Gamma$.  
}
\label{fig3}
\end{figure}

Generally, linear basis sets, \textit{i.e.} an ansatz with linear dependence on the variational parameters, are desirable due to the ease in the evaluation of the gradients, as well as the reduced computational cost of performing simulations with them. The choice of the basis sets can be easily tuned to the symmetries of the system with examples illustrated in Fig.~\ref{fig3}a including Fourier series for periodic potentials, Laguerre polynomials for long-range pairwise interactions, or Jacobi polynomials to represent time-dependence~\cite{das2019variational,das2021reinforcement,das2022direct}. For problems that involve collective dynamics where the dynamically relevant descriptors are challenging to approximate \textit{a-priori}, physics-informed neural-networks offer an alternative ansatz for representing the forces and have been used recently for solving a multitude of different problems within physics that are grounded on stochastic optimal control ~\cite{yan2022learning,boffi2024deep,singh2023variational,holdijk2024stochastic,kang2024computing}. 
Deep models can be advantageous for constructing many-body controllers that can be adapted to the symmetries of the system, however they can be computationally expensive depending on the size of the system and method of implementation.  

For systems with sets of fast and slow variables, such as an underdamped system evolving with a large friction coefficient, methods from homogenization theory \cite{pavliotis2016stochastic} can be utilized to marginalize over the fast variables and express the controller solely in terms of the slow variables \cite{singh2023variational,hartmann2014optimal}. Finally, for the case of finite-time conditioning, learning the optimal controller requires learning both its spatial and temporal features \cite{das2021reinforcement,das2022direct}. The time-dependence can be parameterized based on knowledge from exactly solvable models, which show that the potential varies away from the observation time over timescales associated with its intrinsic relaxation time. For a reactive conditioning this means that the potential is time independent when $k_{AB}(\tau-t)$ is large, for $t<\tau$~\cite{singh2024splitting}.

\subsection{Optimization}
Once the ansatz has been chosen, the next step involves minimization of the variational loss functions to obtain the optimal controller. Similar to the construction of the ansatz, the most effective method of optimization depends on the specific type of problem being studied. In the past, we have adopted two distinct approaches that are best referred to as a direct \emph{fitting} approach, or a reinforcement \emph{learning} approach, both illustrated in Fig.~\ref{fig3}b.

When a conditioned ensemble is accessible via other importance sampling methods, a variational loss can be constructed by reweighting Eqs. \ref{eqvpsk} and \ref{eqvpsldf} back into the undriven conditioned trajectory ensembles  \cite{singh2023variational,rosa2024variational}. Since the control force then only enters into the relative action, not the trajectory probability, this form of optimization can be done directly. The relative action is fit with either a linear or nonlinear basis set. For linear ansatze, the optimization can be accomplished by a simple matrix inversion, and for a nonlinear model, the optimization can be performed through gradient descent.
In terms of applications, this method of optimization is appropriate for finite-time processes where computation and analysis of the commitment probability is of interest \cite{rotskoff2022active,thiede2019galerkin,hasyim2022supervised,jung2023machine,strahan2023predicting,khoo2019solving,liang2023probing}, and amenable to problems in equilibrium where the conditioned path ensemble can be accessed with relative ease by leveraging state-of-the art methods for sampling the reactive path ensembles \cite{jung2023machine}. Within nonequilibrium steady-states however, the applicability of this fitting procedure is diminished by the lack of efficient sampling tools that afford easy access to the conditioned path ensemble.

 For cases where a conditioned ensemble is not available directly, VPS can iteratively generate 
 rare trajectories and evaluate their likelihoods by learning the control force on the fly. 
 The iterative optimization algorithm consists of the choice of initial values of variational parameters, simulation of an ensemble of driven trajectories to compute the variational estimate and its gradients, and updating the parameters for the next iteration with the gradients. At convergence, the optimized force simultaneously gives an estimate of the rate as well as generates rare trajectories as typical ones. Particular advantage can be gained if the force is expressed as a linear combination of a complete set of basis functions. This allows easy differentiation of the force with respect to the variational parameters. This approach efficiently combines the evaluation of forces at every timestep to the accumulation of the gradient estimators for variational optimization. For the infinite time-conditioning, temporal self-averaging in steady-state can be used to improve the estimation of the gradients of the force with respect to the variational parameters. This is accomplished with the use of so-called Malliavin weights~\cite{warren2012malliavin}, which are auxilliary variables propagated along with the real dynamical variables in the system. Each Malliavin weight keeps the gradient of the trajectory probability with respect to a variational parameter. At the end of the steady-state trajectory, the gradient of the cost-function is directly estimated from the correlation function of the Malliavin weights with the observable. 

For complex, interacting systems where the bases sets span more than a few dimensions, the combinatorial explosion of the number of spatiotemporal variational parameters in learning implies that the statistical fluctuations in each cost-function gradient is too large to learn efficiently with a naive gradient descent algorithm. Specialized reinforcement learning tools have been developed for this purpose that learn not only the force, but also a value-function for expected future cost starting from any point in space and time. This approach reduces the variance of the gradient estimate drastically by subtracting a zero-mean high-variance baseline. Such specialized reinforcement learning algorithms such as Monte Carlo with a value baseline and the actor-critic algorithm, can learn an optimal controller sufficiently accurately that the reaction rate, scaled cumulant generating function and reactive trajectory ensemble are quantitatively estimated \cite{das2021reinforcement,das2022direct}. Alternative optimization methods also exist for construction of the effective approximate controllers. These include feedback algorithms~\cite{nemoto2014computation,nemoto2016population}, brute force enumeration of variational parameters~\cite{jacobson2019direct}, or  evolutionary algorithms~\cite{whitelam2020evolutionary}. 
Recently the exact controller for many-particle systems in one and two dimensions have been constructed with tensor network basis sets  \cite{banuls2019using,causer2023optimal}. 
This approach has still been limited to spatially discrete systems but holds promise in its treatment of fluctuations that are many-body in origin. 

\subsection{Estimation of rare event statistics}
The optimal force is in general many-bodied, but a low-rank ansatz is often a tractable approximation. For example, a control force might be truncated at one and two-bodied forces. 
We have found that such an approximate force is  able to describe the principal mechanism leading to rare fluctuations as discussed in the following sections. In such cases where the variational bound is nearly saturated, the approximation error from a limited basis set can be  corrected for by estimating the exponential averages in Eqs. \ref{eqvpsk} and \ref{eqvpsldf} either directly, or through a cumulant expansion \cite{das2019variational}. In case that both the driven and conditioned or tilted path ensembles are available, and the driving force approximation is accurate enough that there is sufficient overlap between the two distributions, as shown in Fig.~\ref{fig3}c, the likelihood of the rare fluctuation can be evaluated using a generalization of the Bennett Acceptance Ratio \cite{das2022direct,singh2024splitting,frenkel2023understanding}.

There are cases where the bound is sufficiently unsaturated that converging the correction is not numerically possible. In this case, the calculation of the exponential averages can also be performed iteratively in small exponential increments with a cloning algorithm \cite{giardina2006direct,ray2018exact}. Using optimized forces from VPS speeds up the calculation by many orders of magnitude and has been shown to outperform such stratification based sampling algorithms such as cloning and Forward Flux Sampling \cite{das2019variational,das2022direct}.  Future work is needed on developing such hybrid efficient techniques in finite-time for calculating reaction rates.

\section{Applications to rate constants} 
VPS has been used to estimate rate constants and characterize rare fluctuations, enabling mechanistic studies of reactive events both in and out of equilibrium. The algorithms in Section 4 offer various means for computation of reactive rates from both conditioned and driven reactive trajectory ensembles, offering robust means for estimation of reactive probabilities. Working in the conditioned ensemble enables computation of the many-body reaction coordinates, and through the framework of stochastic optimal control, enables simple quantitative metrics for distilling rare fluctuations on a microscopic level of detail. 

\subsection{Quantifying mechanisms}
A long-standing goal in theoretical chemistry has been to develop methods to distill dynamics on a high dimensional landscape into a few physically meaningful descriptors 
\cite{peters2017reaction}. Within the  
framework of Transition path sampling \cite{bolhuis2000reaction} and Transition path theory \cite{vanden2010transition} the committor has been identified as the ideal reaction coordinate.
However, as the solution of the Backward Kolmogorov equation, the committor formally lives in the full dimensional space of the system.

\begin{figure}
\includegraphics[width=16cm]{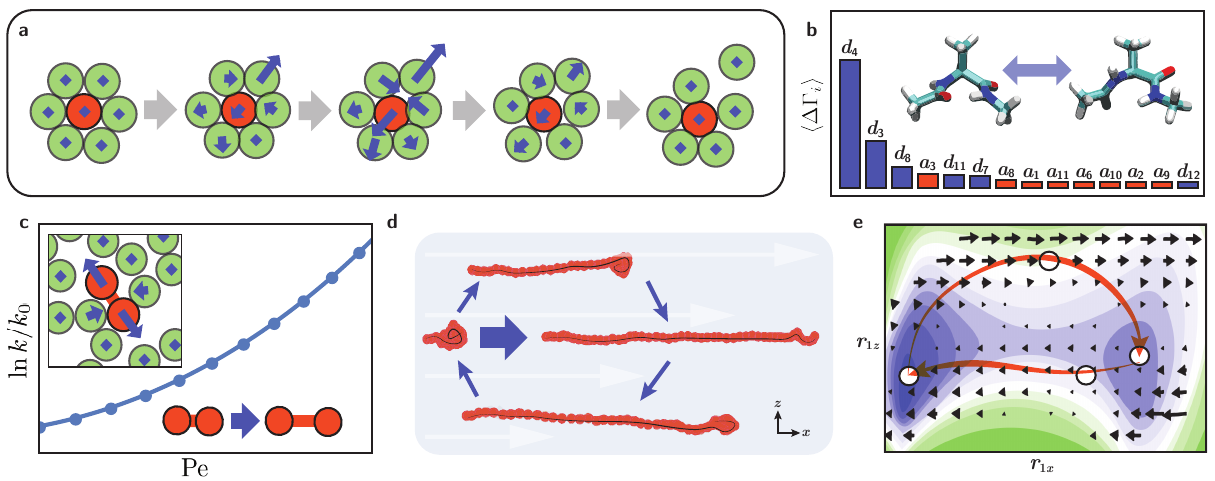}
\caption{Mechanisms of reactive events clarified with VPS. {\textbf a} The optimal control forces (blue arrows) for the dissociation (or association) of a self-assembled colloidal cluster demonstrate a T1 like mechanism. {\textbf b} The relative contributions of each dihedral ($d_i$) and angle ($a_i$) to the isomerization of alanine dipeptide. {\textbf c} Rate relative to its equilibrium value, $k/k_0$ with increasing activity for the conformational change of a passive dimer in an active fluid with optimal control forces in the inset (blue arrows). {\textbf d} The shear induced unfolding of a polymer proceeds through distinct mechanisms in the folding and unfolding pathways due to persistent probability currents as illustrated in {\textbf e} by the projection of the mean current (black arrows) and log-likelihood (contour plot) onto two Rouse modes along the shear gradient $\bf{r}_{1z}$ and direction of the shear $\bf{r}_{1x}$. Typical unfolding (top) and unfolding (bottom) pathways are shown in orange.}
\label{fig4}
\end{figure}
Within this context, the application of theories of stochastic optimal control admit a rather alternative perspective into the analysis of these highly nonlinear functions, which can be summarized as follows: (1) any conditioned Markov process can be mapped onto an unconditioned Markov process with an external drift that encapsulates all the natural fluctuations that induce the conditioning (2) this external drift can be related to the conditional probability itself (3) the natural fluctuations that induce the conditioning can be \textit{quantified} by the computation of the change of measure between systems with and without this external drift, and generally be decomposed into effective contributions from different degrees of freedom. As such, the formalism of VPS \cite{das2019variational,das2022direct,singh2023variational,singh2024splitting,rosa2024variational,singh2024nonequilibrium} and of similar methods based on stochastic optimal control \cite{boffi2024deep,yan2022learning} can be utilized to visualize and quantify rare fluctuations on a \textit{molecular} level of detail.

\subsubsection{Colloidal cluster dissociation}
The simple relation between the optimal time-dependent controller that solves the generalized bridge problem and the splitting probability was employed in Ref. \cite{singh2024splitting} to study the dissociation of a particle from a heptameric cluster, originally inspired from Ref. \cite{dellago1998transition} and used in a range of different studies \cite{das2021variational,das2023nonequilibrium,yuan2024optimal}. The variational form in Refs. \cite{khoo2019solving,rotskoff2022active,hasyim2022supervised} was used to construct the splitting probability and the optimal controller was used to generate reactive trajectories. Fig. \ref{fig4}a shows a snapshot of a representative reactive trajectory with the optimized control forces illustrated by the arrows. These control forces encapsulate the natural fluctuations of the system that make a trajectory reactive, and this system offers an illustrative example of how the formalism of VPS can be employed to visualize many-body fluctuations of rare events that involve collective motion. What emerged in this case is reminiscent of a T1 transition \cite{hasyim2021theory}, a characteristic deformation pattern for switching neighbors in dense media. Utilizing this formalism for studying collective rare events such as nucleation and developing Monte-Carlo moves for importance sampling of glassy systems offer promising avenues of future research.

\subsubsection{Alanine dipeptide isomerization}
For cases where the high dimensionality of the system restricts visualization of the forces, one can utilize the fact that the stochastic action due to the control forces $\langle \Delta \Gamma \rangle$ can be decomposed into effective contributions from different degrees of freedom to infer mechanistic insight. This framework affords a direct quantification of rare fluctuations along different descriptors and can be transformed across different representations of the system, allowing simple ways to perform dynamically consistent coarse-graining along physically relevant descriptors. Its illustration was shown in the study of isomerization of alanine dipeptide that serves as a standard model for exploration of unimolecular reactions in interacting systems \cite{bolhuis2000reaction}.  The reaction was explored by fitting the time-dependent committor \cite{singh2023variational} from trajectories obtained from transition path sampling, and a bond-angle-torsion transformation of the system was used to represent the set of descriptors. Figure \ref{fig4}b shows the conditioned action along the leading descriptors that were identified by the method. The most important descriptor was found to be the Ramachandran angle $\phi$ ($d_4$) consistent with previous studies on the system \cite{bolhuis2000reaction}, and the analysis additionally highlighted the relative motion of the alkyl bond to be activated during the transition.

\subsection{Rate enhancement and stability} 
A key aspect of the discussion in previous sections were that all the methods of analysis do not invoke detailed balance, and can be employed to characterize reactive events in nonequilibrium steady states. In this section, we explore the application of VPS to two examples of reactions far from equilibrium that highlight features of nonequilibrium steady states that are strongly relevant to the endeavors of physical chemistry. 

\subsubsection{Dimer in an active solvent} 
A paradigmatic model for nonequilibrium soft matter is active matter, where energy is continuously injected to keep particles autonomously motile. Due to an intrinsic self-propulsion, active matter often exhibits unusual dynamics such as ballistic motion at short timescales \cite{redner2013structure}, formation of living clusters~\cite{mognetti2013living}, swarming~\cite{vicsek1995novel} and motile topological defects \cite{marchetti2013hydrodynamics}. Such dynamics are common across a range of length-scales in biological systems such as cytoskeleton in cells, colonies of bacteria and flocks of birds \cite{ramaswamy2010mechanics}. Besides being useful models for living systems, active matter can be a tool for designing self-organized materials that show novel structure and function. A general quest has been to find ways to extract useful work from an active system. This is challenging because the directions of self-propulsion in neighboring active particles are not always aligned; thus a large portion of the self-propulsion energy is lost as dissipation through interparticle collisions. A new approach to solve the problem of harvesting energy from active systems has been to design passive solute particles or obstacles that are then driven by their active environment \cite{grunwald2016exploiting,mallory2018active,sokolov2010swimming}. It is in this spirit that VPS was recently applied to quantify the rate and mechanism of isomerization of a colloidal dimer in a dense active fluid~\cite{das2019variational}. As shown in Fig. \ref{fig4}c , the colloidal dimer has an extended and collapsed state. When it is placed in a bath of active Brownian particles and the self-propulsion of the bath is gradually increased, the rate of reaction of the dimer from the collapsed to the extended state is enhanced quadratically by a factor of twenty. The mechanism of the enhancement was revealed from the most probable reactive trajectories extracted from those driven with the optimized force. The mechanism is shown in the inset of Fig. \ref{fig4}c and illustrates that a long-lived collective polarization fluctuation of the active fluid is responsible for sterically driving the extension reaction. This observation is in line with prior experiments on active particles adhering to passive surface depending of the surface geometry \cite{sokolov2010swimming} and demonstrated the ability of VPS to estimate rates and quantify many-body reaction mechanisms in nonequilibrium matter.

\subsubsection{Attractive polymer in a shear flow}
Emergence of metastability of thermodynamically unfavorable states is often observed in systems that are driven out of equilibrium. The exploration of these systems is an active area of research in active matter \cite{marchetti2013hydrodynamics, cates2015motility}, self-assembly \cite{mallory2018active,nguyen2021organization,palacci2013living} and biophysics \cite{alexander2006shear,tan2022odd}. In the case that the emergent state is metastable, inference into the mechanism of the reaction is challenging due to the existence of persistent probability currents that can mediate reactions~\cite{fang2019nonequilibrium,das2023nonequilibrium}. Standard approaches have leveraged effective equilibrium based theories~\cite{redner2016classical,cates2023classical}, extensions of transition path theory to time-irreversible systems \cite{helfmann2020extending} and computation of minimum action paths \cite{zakine2023minimum,heller2024evaluation} to infer mechanisms into reactions in nonequilibrium steady states. 

An example of this phenomenology is a grafted polymer with strong attractive interactions under shear, that has been employed to gain insight into the role of von Willebrand cofactor in the process of blood clotting \cite{alexander2006shear,schneider2007shear}. These polymers have been observed to exhibit a globule-stretch transition at a critical shear as shown in Fig. \ref{fig4}d, similar to the unfolding of the  von Willebrand cofactor into thin fibers under physiologically relevant shear rates. Using the formalism of VPS and leveraging the transferability of the metric for discovering relevant dynamical coordinates, the leading normal modes $r$ were identified as the best set of descriptors that were activated during globule-stretch transitions. Since the stretched state was  thermodynamically unstable in equilibrium, it was not surprising that the nonequilibrium reaction coordinates were found to be orthogonal to the equilibrium reaction coordinate. Moreover, due to the breakdown of detailed balance, the forward and backward transition in the nonequilibrium steady state were found to be distinct from each other, with the pathways passing through different intermediates as shown in Fig. \ref{fig4}d. Both of these features can be visualized in Fig. \ref{fig4}e that shows the negative log-likelihood and the net flux at finite shear along the $ r_{1x}$ and $r_{1z}$, Rouse modes that correspond to the equilibrium and one of the nonequilibrium reaction coordinate respectively, along with representative folding and unfolding reactive pathways. 
The system illustrates how probability currents can mediate reactions. Instead of a standard `barrier crossing' transition, the unfolding reaction involves coupling to strong unidirectional currents observed at the top of Fig. \ref{fig4}e. Similarly, the folding reaction has a low free energy barrier that can be overcome through thermal fluctuations, and requires escape from the strong cyclic currents that confine the polymer within the extended state. 

\section{Applications to large deviation functions} 
Both the rate function and the scaled cumulant generating function, as defined through Eqs.~\ref{eqratef} and \ref{scgf}, emerge as the solutions to several variational principles. 
Variational principles exist because large deviations are achieved in the \textit{least rare} of all rare pathways. This idea is called the contraction principle for large deviations~\cite{chetrite2015variational}. 
The contraction principle underpins the expressions for large deviation functions in terms of control forces that can achieve a given large deviation, such as for the scaled cumulant generating function in Eq.~\ref{eqvpsldf}.
The saturation of Jensen's inequality with the control force $-\nabla \Lambda^{*}$, the solution of the Hamilton-Jacobi-Bellman equation, Eq.~\ref{eqHJB}, provides the mechanism of this least rare set of fluctuations. Solving for the optimal force thus gives us quantitative estimates for the rate functions and the mechanism of the rare event.

\begin{figure}
\includegraphics[width=16cm]{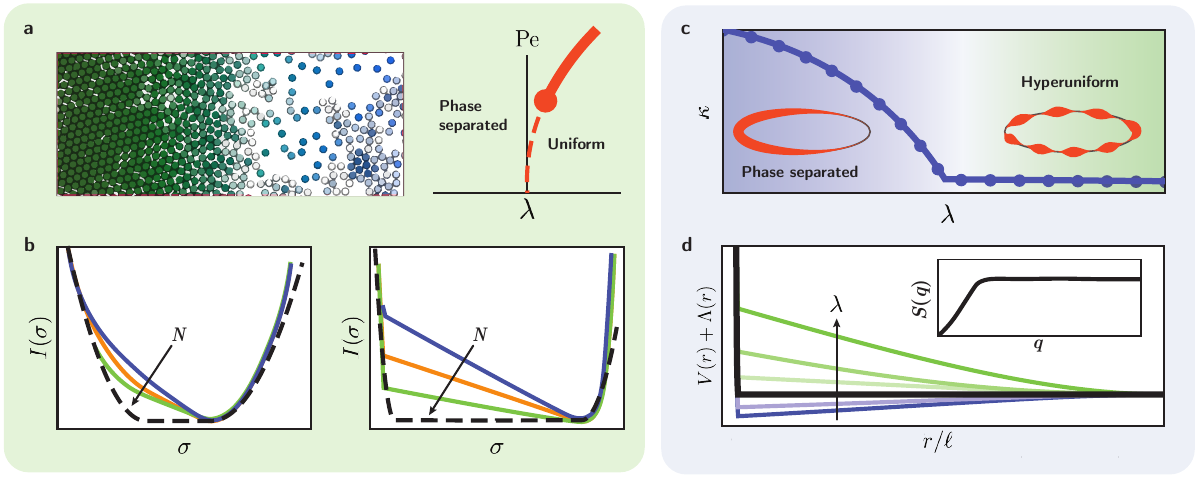}
\caption{Large deviation function calculations with VPS include studies on {\textbf a} phase separation in active fluids, whose diagram depends on activity (Pe) and tilting parameter ($\lambda$). {\textbf b} Finite size scaling analysis of the entropy production rate functions with particle number $N$ were computed from VPS (solid lines), with theoretical calculations yielding an asymptotic form (dashed lines) at low (left) and high (right) Pe numbers. {\textbf c} Soft Brownian particles on a line exhibit a dynamical phase transition as a function of tilting parameter between phase separated and hyperuniform states, stabilized by control forces shown in {\textbf d}  whose range was much longer then the particle size ($\ell$). (inset) Hyperuniformity is defined as the suppression of density fluctuations, evident in the behavior of the structure factor, $S(q)$ at low wavevectors $q$.}
\label{fig5}
\end{figure}
\subsection{Dynamic phase transitions} 
When the rare event involves the coexistence of distinct macroscopic dynamics for an extensive amount of time, such as formation of a macroscopic cluster starting from a dispersed phase, it is said to be in a different dynamical phase. Transitions between dynamical phases with different values of a dynamical observable are associated with bistability in $I(\mathcal{O}/\tau)$ or  singularities in derivatives of $\ln Z_\lambda$~\cite{buvca2019exact,garrahan2007dynamical,espigares2013dynamical,jack2015effective}. For example, a discontinuity in the first derivative of the $\ln Z_\lambda$ for the current denotes the coexistence of distinct phases that have different mean values of current. 

A classic example of a dynamic phase transition occurs in active matter.
Amongst the diverse phenomenology exhibited by active matter~\cite{ramaswamy2010mechanics}, the emergence of stable phase separation in purely repulsive particles as shown in Fig.~\ref{fig5}a garnered significant theoretical attention. The phase diagram describing the so-called motility induced phase separation was worked out as a function of Pe number, a measure of active to diffusive motions, using theoretical arguments and direct simulations~\cite{cates2015motility}. The entropy production was found to be a useful order-parameter distinguishing the condensed from dilute phase, and controlling the the entropy production could trigger phase separation~\cite{nemoto2019optimizing,fodor2022irreversibility}, at least for linear theories~\cite{tociu2019dissipation}. In Ref.~\cite{grandpre2021entropy} a general lower bound on distributions of entropy production for active matter systems was derived using the variational principles underpinning VPS. An approximate control force was used to sample entropy production fluctuations consisting of the exact solution to the single particle problem and using a linearization of a hydrodynamic model motivated from work on hard rods~\cite{dolezal2019large}. It was found that for systems far away from dynamical phase transitions, the bound was tight, and its weakening signaled the importance of long-ranged correlations. This observation linked the locations of phase transitions to enhanced entropy production fluctuations as probed by a tilted ensemble, shown in Fig.~\ref{fig5}a. Large deviation theory and finite size scaling of the rate function was used to motivate a generalization of a Maxwell construction between the two phases, Fig.~\ref{fig5}b. 

\subsection{Long-ranged forces} 
The steady-state of many-particle systems out of equilibrium in general has long-range correlations \cite{spohn1983long}. Even for dynamics that does not lead to long-range correlations in their typical behavior, rare fluctuations of size-extensive variables in general result from long-range correlations between particles. The corresponding optimal control forces are long-ranged, often in the form of generalized Coulombic interactions \cite{popkov2010asep}. While analytical approximations for the optimal control forces can be obtained for small biases in the hydrodynamic limit using macroscopic fluctuation theory \cite{bertini2015macroscopic}, they are not available for complicated interparticle interactions. VPS can used to numerically learn the optimal forces provided a long-range ansatz is used in the basis functions. 

VPS was applied to a model of overdamped repulsive particles in one dimension to characterize fluctuations of dynamical activity ($\kappa$). Dynamical activity measures the time-integrated propensity of particles to move and explore the space around them \cite{pitard2011dynamic}. Activity has been used as a dynamical order parameter to study the glassiness of materials \cite{hedges2009dynamic}, the frequency of dissipative transitions in a model of trapped ions \cite{ates2012dynamical}, and to characterize fluids that show long-range correlations despite disorder \cite{jack2015effective,torquato2018hyperuniform}. The last example, called a hyperuniform phase, arises in the one-dimensional model for low magnitudes of dynamical activity fluctuations. Similar models have been shown to have a dynamical phase transition at near-zero tilting, $\lambda \approx 0$, crossing between a hyperuniform phase and a clustered state with macroscopic phase separation~\cite{jack2015effective,dolezal2019large}. Using macroscopic fluctuation theory, the hyperuniform phase can be shown to exhibit pairwise long-range repulsive interactions that push the particles away from each other so that collisions are minimized. 

VPS enabled the quantification of the two phases outside of the regimes of validity of approximate theory~\cite{das2019variational}. When the ansatz for learning the optimal force allows for long-range interactions via  basis functions of scaled Laguerre polynomials, VPS learns $(\ln Z_\lambda)/\tau$ and the correct rare trajectories leading to high values of activity (phase-separated) and low values (hyperuniform), as shown in Fig. \ref{fig5}c. The clustered phase in this case has a quadratic scaling of the activity with respect to system size, as the most long-lived clustered configurations are achieved by minimizing collisions on the surface of the cluster as opposed to in its center, where particles have collisions with both neighboring particles at once. In contrast, the hyperuniform phase consists of particles colliding only pairwise at a time and has a linearly scaling activity. The optimal force here consists of a long-range repulsion that increases strongly with the bias, and leads to a suppression of long-wavelength density fluctuations (structure factor $S(q)\to0$ as $q\to0$), as shown in Fig. \ref{fig5}d.

\section{Future directions}
The preceding sections describe applications from several branches of chemical, material and biological physics. The approaches we have reviewed here allow for the calculation of rates, and the evaluation of large deviation functions for high dimensional, complex systems evolving arbitrarily away from equilibrium. However, there is still much work to be done to understand the structure of the theory underpinning variational path sampling. The structure of the variational landscape in terms of control forces and its relation to the relaxation timescales of the system may affect the convergence of the variational algorithm; yet this behavior is not formally well-understood. Further, the application of VPS to some systems is no doubt still difficult owing to large statistical uncertainties in the formally computed parameter gradients. Understanding the best way to construct control forces spanned by low dimensional collective coordinates is unclear. Improvements in their accuracy and more knowledge of the best statistical estimators to use in these cases would help in extending VPS to processes that are not spatially local like nucleation. Hybrid approaches in which optimized control forces are added to transition path sampling or the cloning algorithm~\cite{das2019variational,das2022direct}  have shown promise and warrant continued study. Extending these approaches to other methods like nonequilibrium instantons~\cite{heller2024evaluation}, or nonequilibrium umbrella sampling~\cite{dickson2010enhanced} would likely also be fruitful. Finally, extending VPS to the evaluation of other observables like stationary distribution functions or multi-time functions would increase the scope of problems tractable within its application. Early work along the former direction has just begun~\cite{rosa2024variational}. It is our hope that others’ experience and diverse perspectives will lead to advances in these areas and unanticipated directions.

\section*{DISCLOSURE STATEMENT}
The authors are not aware of any affiliations, memberships, funding, or financial holdings that might be perceived as affecting the objectivity of this review. 

\section*{ACKNOWLEDGMENTS}
The work reviewed in this article was supported largely by NSF Grant CHE1954580 and U.S. Department of Energy, Office of Science, Office of Advanced Scientific Computing Research, and Office of Basic Energy Sciences, via the SciDAC program. Several members of the Limmer Group and other colleagues have contributed significantly to the development of the ideas described in this review including Phillip Geissler, Juan Garrahan, Garnet Chan, Robert Jack, Benjamin Rotenberg, Dominic Rose, Ben Kuznets-Speck, Chloe Gao, Trevor GrandPre, Anthony Poggioli, Ushnish Ray, Jorge Rosa-Raíces, and Eric J. Heller.

\bibliography{final-vps-arpc}

\end{document}